\documentclass[compsoc,conference,a4paper,10pt,times]{IEEEtran}
\IEEEoverridecommandlockouts

\usepackage[colorlinks=true,urlcolor=black]{hyperref}
\usepackage{float}
\restylefloat{figure}
\restylefloat{table}
\usepackage{graphicx} 
\usepackage{amsmath}
\usepackage{url}
\usepackage{listings}
\usepackage{wasysym}
\usepackage{enumitem}
\usepackage{booktabs}
\usepackage{caption}
\usepackage[table,xcdraw]{xcolor}
\usepackage{soul,xcolor}
\usepackage[numbers]{natbib}

\usepackage{scalerel}
\usepackage{tikz}
\usetikzlibrary{svg.path}

\definecolor{orcidlogocol}{HTML}{A6CE39}
\tikzset{
  orcidlogo/.pic={
    \fill[orcidlogocol] svg{M256,128c0,70.7-57.3,128-128,128C57.3,256,0,198.7,0,128C0,57.3,57.3,0,128,0C198.7,0,256,57.3,256,128z};
    \fill[white] svg{M86.3,186.2H70.9V79.1h15.4v48.4V186.2z}
                 svg{M108.9,79.1h41.6c39.6,0,57,28.3,57,53.6c0,27.5-21.5,53.6-56.8,53.6h-41.8V79.1z M124.3,172.4h24.5c34.9,0,42.9-26.5,42.9-39.7c0-21.5-13.7-39.7-43.7-39.7h-23.7V172.4z}
                 svg{M88.7,56.8c0,5.5-4.5,10.1-10.1,10.1c-5.6,0-10.1-4.6-10.1-10.1c0-5.6,4.5-10.1,10.1-10.1C84.2,46.7,88.7,51.3,88.7,56.8z};
  }
}

\newcommand\orcidicon[1]{\href{https://orcid.org/#1}{\mbox{\scalerel*{
\begin{tikzpicture}[yscale=-1,transform shape]
\pic{orcidlogo};
\end{tikzpicture}
}{|}}}}

\begin{document}

\title{SafePickle: Robust and Generic ML Detection of Malicious Pickle-based ML Models}


\author{
\IEEEauthorblockN{Hillel Ouri Ohayon \orcidicon{0009-0001-6598-638X}}
\IEEEauthorblockA{\textit{Department of Computer and Software Engineering} \\ \textit{Ariel Cyber Innovation Center}\\ Ariel University, Israel\\
hillel.ohayon@msmail.ariel.ac.il}
\and
\IEEEauthorblockN{Daniel Gilkarov \orcidicon{0009-0008-9274-802X}}
\IEEEauthorblockA{\textit{Department of Computer and Software Engineering} \\ \textit{Ariel Cyber Innovation Center}\\ Ariel University, Israel\\
daniel.gilkarov1@msmail.ariel.ac.il}
\and
\IEEEauthorblockN{Ran Dubin \orcidicon{0000-0002-2055-2211}}
\IEEEauthorblockA{\textit{Department of Computer and Software Engineering} \\ \textit{Ariel Cyber Innovation Center}\\ Ariel University, Israel\\
rand@ariel.ac.il}
}

\maketitle



\begin{abstract}
Model repositories such as Hugging Face increasingly distribute machine learning artifacts serialized with Python’s \texttt{pickle} format, exposing users to remote code execution (RCE) risks during model loading. 
Recent defenses, such as PickleBall, rely on per-library policy synthesis that requires complex system setups and verified benign models, which limits scalability and generalization. 
In this work, we propose a lightweight, \textbf{machine-learning-based scanner} that detects malicious Pickle-based files without policy generation or code instrumentation. 
Our approach statically extracts structural and semantic features from Pickle bytecode and applies supervised and unsupervised models to classify files as benign or malicious. 
We construct and release a labeled dataset of 727 Pickle-based files from Hugging Face and evaluate our models on four datasets: our own, PickleBall (out-of-distribution), Hide-and-Seek (9 advanced evasive malicious models), and synthetic \texttt{joblib} files. 
Our method achieves \textbf{90.01\% F1-score} compared with 7.23\%-62.75\% achieved by the SOTA scanners (Modelscan, Fickling, ClamAV, VirusTotal) on our dataset.
Furthermore, on the PickleBall data (OOD), it achieves \textbf{81.22\% F1-score} compared with 76.09\% achieved by the PickleBall method, while remaining fully library-agnostic. 
Finally, we show that our method is the only one to correctly parse and classify 9/9 evasive Hide-and-Seek malicious models specially crafted to evade scanners.
This demonstrates that data-driven detection can effectively and generically mitigate Pickle-based model file attacks.

\end{abstract}

\section{Introduction}
\label{introduction}

Pre-trained Model (PTM) sharing \cite{davis2023reusing} is a key driver of the wide adoption of AI among researchers, developers, and laypeople alike. People can leverage the rich variety of PTMs to use AI without the costly requirement of training a model from scratch.
Nowadays, techniques like Low-Rank Adaptation (LoRa) \cite{lora} allow laymen with very limited resources to create their own customized fine-tuned models. This clearly makes AI more accessible than ever before; evidently, for example, at the time of writing this article, the \href{https://Hugging Face.co/black-forest-labs/FLUX.1-dev}{FLUX.1-dev} image generation generative AI model on the Hugging Face PTM sharing hub alone has over 36k adapters made by users.
PTM sharing hubs such as Hugging Face \cite{Huggingfacecite} are immensely popular, currently hosting almost 2.2M model repositories, with billions of downloads per month \cite{jiang2023empiricalstudypretrainedmodel,ptm_hf_survey}.


Unfortunately, while PTM sharing has great benefits, it also introduces security challenges that can serve cybercriminals in malicious activity \cite{mitre_atlas}, much like the ones seen in open software sharing and the software supply chain \cite{ohm2020backstabber}.
One major attack pattern is Remote Code Execution (RCE) during model deserialization: attackers leverage vulnerabilities in the serialization formats of common ML/AI libraries such as: PyTorch \cite{Pickle_serialization_vul,pytorch_serialization}, TensorFlow \cite{tf_abuse}, and Scikit-Learn \cite{malhug} to craft malicious models that that they can then easily propagate via PTM sharing hubs such as Hugging Face.
The most common serialization format is Python's Pickle \cite{pickle_standard}, which is very commonly used, for example, by PyTorch and Scikit-Learn, two very popular ML libraries.
In this paper, we focus on detecting Pickle-based malicious model attacks.

The Python Pickle module is widely known to have RCE vulnerabilities, dating back as far as 2011 \cite{Pickle_serialization_vul}. It allows an attacker to execute any Python function they wish once the malicious Pickle-based file is loaded (e.g., PyTorch, Joblib, etc.). Despite this, and despite the existence of safer alternatives such as Hugging Face's Safetensors \cite{Safetensors} file format, especially created to alleviate this type of threat, we can observe that in practice, users on Hugging Face are still choosing to use Pickle \cite{PickleBall2025}. 
Moreover, while the RCE threat is also found in other file formats, such as the one used by TensorFlow \cite{tf_abuse}, 95\% of the malicious models observed on Hugging Face are Pickle-based (PyTorch) \cite{jfrog_article1}, hence, we especially focus on directly treating the Pickle-based file threat.

Previous attempts to mitigate the Pickle-based file threat can be grouped into 3 categories:

\noindent \textbf{(1) Model Scanners:} these methods are used to analyze the Pickle content, and flag it as safe or unsafe. This approach is used by Hugging Face, by executing various scanners on each file uploaded, requiring the user to make a decision whether to trust the file or not given the scan results. In practice, currently existing scanners mostly use the same shallow approach of just looking at the imports used within the Pickle file. These scanners often label benign models as malicious \cite{malhug,PickleBall2025}, partially since they're coarse-grained classifiers.
In this work, we introduce a novel ML-based model scanner that uses deeper features to achieve more robust results.

\noindent \textbf{(2) Safe-loading:} These methods aim to replace the regular, dangerous Pickle loading mechanism with a safer alternative. 
Examples include PyTorch's weights-only unpickler, and PickleBall \cite{PickleBall2025}.
Compared with scanning (that doesn't require execution of the Pickle), these methods might be dangerous, as they're loading a file that might be harmful and that might bypass the security measures, as outlined by the Hide-and-Seek \cite{hide_and_seek} paper.
Moreover, safe-loaders are shown to disrupt common legitimate workflows (PyTorch's safe-loader prevents 15\% of files on Hugging Face from loading \cite{PickleBall2025}), or they require the user to manually create a loading policy fit for their situation \cite{PickleBall2025}, which severely limits the approach.  

\noindent \textbf{(3) Replacing Pickle with another serialization format:} This approach aims to replace Pickle entirely with a safer serialization format such as Safetensors. In practice, we can see that this approach is not embraced by AI users, who continue to choose Pickle despite its drawbacks due to its convenience, flexibility, etc.

To address these limitations, we present \textbf{SafePickle}, a generic drop-in ML detection system that classifies Pickle-based files based on extracted features.
This approach is shown to lower false-positive and false-negative results, commonly plaguing existing scanners that Hugging Face uses \cite{malhug}, and it is much simpler to use compared with safe-loading approaches such as PickleBall \cite{PickleBall2025}, which requires creating a policy for each library the user wishes to use.

We evaluated and compared \textbf{SafePickle}, and the state-of-the-art methods (scanners, safe-loaders) on 4 datasets: our own dataset that contains 727 models (648 benign, 79 malicious) that we collected off of Hugging Face, the PickleBall \cite{PickleBall2025} dataset, a specially curated dataset of 12 unique malicious model samples that current methods often fail on, and finally, the evasive Hide-and-Seek \cite{hide_and_seek} malicious samples. Our evaluation shows that our approach preserves and improves the false-negative and false-positive results of the state-of-the-art scanners, it achieves similar results to safe-loading methods such as PickleBall \cite{PickleBall2025} while being vastly simpler and generic, and finally, ours is the only method that successfully processes and detects the specially crafted evasive malicious samples from Hide-and-Seek.

\section{Contributions}
Our work proposes a data-driven alternative to policy-based Pickle defenses such as PickleBall. 
Instead of generating per-library policies, we train ML models on extracted Pickle features for generic RCE detection.
The main contributions are:

\begin{itemize}
    \item \textbf{ML-based Pickle malware detection:} 
    We present the first supervised and unsupervised learning approach for detecting malicious Pickle files via static feature extraction, requiring no library-specific policies or benign reference models.

    \item \textbf{Analysis of prior defenses:} 
    We show that (i) restricted unpickling removes functionality without full protection, 
    (ii) Hugging Face scanners suffer extreme FP/FN imbalance, and 
    (iii) PickleBall’s policy synthesis is non-scalable and depends on verified benign models.

    \item \textbf{Open-source dataset:} 
    We release a manually labeled dataset of more than 700 benign/malicious Pickle files from Hugging Face \cite{ourdataset}.

    \item \textbf{Cross-dataset evaluation:} 
    Our models are evaluated on four datasets: ours, PickleBall (OOD), Hide-and-Seek, and synthetic joblib files - showing strong generalization and robustness.

    \item \textbf{Improved detection results:}
    Our best model achieves \textbf{90.01\% F1-Score} on our dataset compared with 7.23\%-62.75\% achieved by the SOTA scanners, and \textbf{81.22\%} on the PickleBall (OOD) dataset, compared with 76.09\% achieved by the PickleBall method, while remaining fully library-agnostic. 
    Finally, it correctly detects all 9/9 evasive malicious models from Hide-and-Seek \cite{hide_and_seek} which SOTA scanners fail to scan.


\end{itemize}

\section{Background}
In this section, we provide background information, including terminology, and the main concepts used throughout the paper.
\subsection{Pickle Serialization}
\label{sec:pickle_ser}
Serialization is the act of saving a program object into persistent storage as a file. The file can then be transferred and loaded later using deserialization (the inverse operation). Popular examples of serialization methods are Comma-Separated Values (CSV), JavaScript Object Notation (JSON), and Extensible Markup Language (XML).
Python's \lstinline{pickle} \cite{pickle_standard} is the standard module for serialization of Python objects. It is very commonly used by Python users to freeze object states mid-run and exchange them with one another.
"pickling" and "unpickling" are terms that are used to mean serialization/deserialization using Pickle, \lstinline{pickle.Pickler} and \lstinline{pickle.Unpickler} are the classes that perform the respective actions, along with \lstinline{pickle.dump/s} and \lstinline{pickle.load/s}.
See Figure \ref{fig:pickle_example} for a simple illustration of saving/loading using Pickle.
\begin{figure}
    \centering
\begin{lstlisting}[language=Python]
    >>> import pickle
    >>> o = "Python is cool :)"
    >>> o_pickled = pickle.dumps(o)
    >>> print(pickle.loads(o_pickled))
    Python is cool :)
\end{lstlisting}
    \caption{Simple Python snippet that saves/loads a Python object using Pickle.}
    \label{fig:pickle_example}
\end{figure}
Pickle is widely used by Python users for exchanging serialized data; Hugging Face (a model sharing hub) alone has more than 1.5M Pickle-based models \cite{malhug,PickleBall2025}.

\subsubsection{The Pickle Mechanism}
Pickle defines a custom bytecode language that consists of OPCODES and a stack. The Unpickler processes the pickle bytestream by decoding an OPCODE from the top of the stack one at a time, and dispatching matching actions in Python. There are currently 68 OPCODES; see \cite{pickleopcode} for an exhaustive list.
OPCODES either define primitive objects (e.g., STRING, LIST, etc.) or actions to perform (e.g., APPEND: "Append an object to a list."). Pickle is designed in a way that allows it to serialize any Python objects, including custom ones; this makes it very useful, but also allows execution of any Python code when loading an object.
\subsubsection{Pickle Allows RCE By Design}
Pickle, and more broadly, Python, is designed with flexibility in mind, which is one of the properties that helped propel Python to being one of the most popular programming languages; however, it also inherently makes it easy to manipulate, with no security checks and balances in place (for example, you can easily override vital objects: \lstinline{open = None}); Pickle allows Remote Code Execution (RCE) by design using via 6 OPCODES (GLOBAL, INST, NEWOBJ, NEWOBJ\_EX, OBJ, REDUCE). The threat is caused either by a Pickle importing a dangerous module (Python modules execute root-level code written in them on import), a dangerous class instantiated on import (malicious code in \_\_new\_\_ or \_\_init\_\_ class methods), or by a function indicated by the Pickle creator to be executed by setting a \_\_reduce\_\_ function on a custom serialized class.

\begin{figure*}[hbtp!]
    \centering
    \includegraphics[width=1\linewidth]{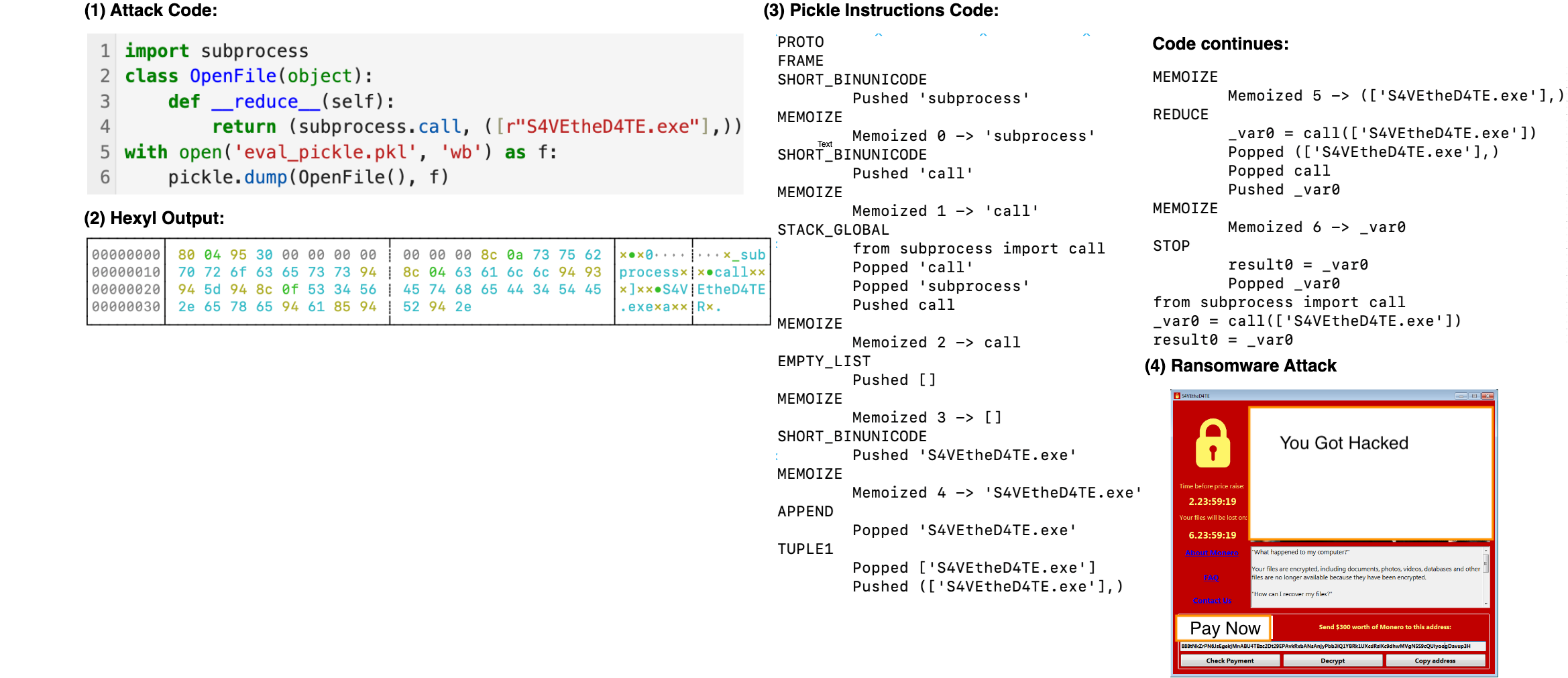}
    \caption{Visualization of Pickle model remote code execution. Step (1) illustrates how easy it is to create the attack, Step (2) is the hex view, Step (3) visualizes how Fickling prints the attack steps, and Step (4) is the execution of the ransomware.}
    \label{fig:pickle_ransom}
\end{figure*}

See Figure \ref{fig:pickle_ransom} for an illustration of how simple it is for an attacker to create a malicious Pickle that causes RCE. This can be done in a few lines using the Python \_\_reduce\_\_ function as seen in Figure \ref{fig:pickle_ransom} step one, lines three and four. Once the model is loaded into memory, the subprocess call will execute the ransomware executable \cite{S4VEtheD4TE} as seen in Figure \ref{fig:pickle_ransom} step four, and the returned value from pickle.load will be the result of the subprocess call. If we examine the raw data of the model using a hex viewer, we can observe that the subprocess call can be seen. However, not all data scientists may be familiar with the validation process and be aware of such details. Some software packages may automatically retrieve the model from a remote repository without notifying the user.

RCE is an immediate threat to unsuspecting victims, as in this case, this allows an attacker to execute any Python program they wish on the victims' machine, and Python even allows executing shell subprocesses, which essentially means an attacker can execute any command they wish with the same elevated privileges the Python program has.
We call this the Pickle Deserialization RCE Exploit (PDRE).

\subsubsection{Restricting Unpicklers}
\label{sec:restricting_unpicklers}
Python suggests a simple patch to address the PDRE by controlling what imports are allowed to happen in an unpickling process by defining a custom Unpickler class.

\begin{figure}[hbtp!]
    \centering
\begin{lstlisting}[language=Python]
import builtins
import io
import pickle

safe_builtins = {
    'range',
    'complex',
    'set',
    'frozenset',
    'slice',
}

class RestrictedUnpickler(pickle.Unpickler):

    def find_class(self, module, name):
        # Only allow safe classes from builtins.
        if module == "builtins" and name in safe_builtins:
            return getattr(builtins, name)
        # Forbid everything else.
        raise pickle.UnpicklingError("global '%s.%s' is forbidden" % (module, name))

def restricted_loads(s):
    """Helper function analogous to pickle.loads()."""
    return RestrictedUnpickler(io.BytesIO(s)).load()
        \end{lstlisting}
\caption{Example of a Restricting Unpickler that only allows importing whitelisted functions.}
\label{fig:runp}
\end{figure}

See Figure \ref{fig:runp} for an illustration; the Unpickler controls what imports are allowed by overloading the find\_class method.
While this might have a minor positive impact, we argue that this approach by itself is insufficient. The Restricting Unpickler blocks calls at the API level, which doesn't allow fine-grained control, and might be either too permissive or too restrictive. For example, the function \lstinline{builtins.getattr} is a threat at the hands of an attacker, but it also has a wide legitimate use-case; for example, the \lstinline{transformers} library automatically serialize \lstinline{training_args.bin} files that use this function, and they are very common. Using this approach, a false-positive (falsely blocking a call) will disrupt the users' workflow, which is sure to cause frustration, and a false-negative (missing an actual attack) will cause potentially catastrophic RCE. \textbf{In conclusion, Restricting Unpicklers are a step in the right direction, but they have devastating limitations that make them unappealing}.
\subsection{AI Model Serialization}
\label{Pickle_Serialization}
Before the distribution or deployment of a trained machine learning (ML) model, the model needs to be serialized.
Different methods exist for serializing ML models and DL models. Pickle is widely used by different ML/DL libraries for serializing trained models; including PyTorch \cite{pytorch_serialization}, Scikit-Learn, etc. These models serialized using Pickle are therefore a potential attack vector, since attackers can leverage the PDRE (see Section \ref{sec:pickle_ser}). 

PyTorch uses Pickle internally to save modules \cite{pytorch_serialization}, tensors, and other PyTorch objects. The library defines the functions torch.save/torch.load for these purposes. A PyTorch saved model in the most up-to-date ZIP file format contains a "data.pkl" Pickle file, and some more files or metadata.
Seeing as PyTorch is one of the most popular AI model Python libraries alongside TensorFlow, in this paper, we focus on them, but the PDRE is possible whenever Pickle is used.
PyTorch officially instructs about serializing models in two ways \cite{pytorch_serialization}: saving the whole model (architecture+weights), this option uses Pickle; and saving just the weights (recommended by PyTorch), this has no PDRE danger. When using the \lstinline{torch.load} function, it requires the user to choose to load potentially unsafe files by setting \lstinline{weights_only=False}.
Generally speaking, choosing to serialize just the weights is advisable in terms of security, however, doing so is either not always possible, or it means the party that wishes to load the file they got has to have the model's architecture code separately, which is inconvenient and makes the model sharing process less self-contained as it is with transferring the whole model in one go (architecture+weights). Hence, this is a utility/security tradeoff.
\label{sec:safetensors}
Attempts have been made to introduce a safe-by-design framework for model sharing, to alleviate the risk associated with Pickle by replacing it; for example, the \textbf{Safetensors} \cite{Safetensors} format introduced by the Hugging Face \cite{Huggingfacecite} model sharing hub. This framework provides a verifiably secure format; however, it only allows for transferring the weights, and the architecture has to be shared separately.
Pickle is likely to remain a staple among users, as it is native to Python, very flexible since it allows for serializing custom objects, and deeply ingrained in existing libraries. This is directly evident from the statistics gathered from Hugging Face \cite{malhug,PickleBall2025}. In the last two yearly quarters, more than 1.3M Pickle-based files were uploaded to Hugging Face model repositories, compared with about 900k Safetensors files.

\subsection{Model Sharing Hubs}
Model sharing hubs are platforms that allow AI users to upload their models to a repository, similar to how GitHub \cite{github} does for software. On the model sharing hubs, users exchange ML/DL models they've trained, discuss them, and can build on top of one another with fine-tuned versions, similarly to repository forks on GitHub. These model sharing hubs have an enormous positive impact on the AI community, helping foster collaboration at an incredible pace, increase accessibility and pushing for the democratization of AI. Various model sharing hubs exist: PyTorch-hub \cite{pytorchhub}, ONNX model zoo \cite{onnx}, Hugging Face (HF) \cite{Huggingfacecite}, and more.
In this paper, we focus on Hugging Face, it has the largest community \cite{malhug}, with over 1.5M model repositories (compared with 1.7k on ONNX model zoo, and less than 100 on PyTorch-Hub), it is library-agnostic, free, and open to use.
Hugging Face allows users to create Model repositories that host one or more ML models in them, and optionally allows for very easy loading of models from the Hugging Face hub using their unifying \lstinline{transformers} Python API that abstracts the underlying model library (TensorFlow, PyTorch, etc.). See Figure \ref{fig:hf_example} for an example of loading a pre-trained model hosted on a Hugging Face repository.

\begin{figure*}
    \centering
\begin{lstlisting}[language=Python]
    # Load model directly
    from transformers import AutoImageProcessor
    from transformers import AutoModelForImageClassification
    
    processor = AutoImageProcessor.from_pretrained(<hf_repo_name>)
    model = AutoModelForImageClassification.from_pretrained(<hf_repo_name>)
\end{lstlisting}
    \caption{Simple Python snippet that loads a pre-trained model from HuggingFace using the transformers Python API.}
    \label{fig:hf_example}
\end{figure*}

\subsection{Pickle API Scanners}
\label{sec:pickle_api_scanners}
Pickle API scanners are programs that scan Pickle files and aim to detect suspicious API calls within them, to help users avoid the Pickle deserialization RCE exploit.
This is the main approach taken nowadays by the available Pickle file scanners, as far as we know.
Generally, they work by scanning the Pickle files for \lstinline{GLOBAL} opcodes that import modules and functions, and flagging them safe/suspicious/unsafe based on a predetermined safe/unsafe import list (this can be seen in the Picklescan source code).
Hugging Face uses the following scanners on every uploaded file: Picklescan \cite{picklescan}, an open-source scanner used by Hugging Face, ProtectAI Modelscan \cite{modelscan}, an open-source ML model scanner that includes Pickle scanning, and finally, JFrog model security scanner \cite{jfrog_ai_model}, a proprietary scanner. There is also Fickling \cite{Fickling}, an open-source Pickle analysis tool.
Picklescan is well-integrated within Hugging Face, it shows a report of detected Pickle imports for each scanned file; the rest of the scanners only show a safe/unsafe result. 
While these scanners help detect well-known attacks, they are shown to miss attacks \cite{pickle-attack}.

These scanners share a common flaw - since they operate at the API level, which only shows what function is used, and not how, scan results often label legitimate Pickle files as unsafe or suspicious.
For example, the Python function \lstinline{builtins.getattr} can be used to gain access to object attributes; this can potentially be useful for an attacker, however, this function is also used regularly for benign uses. The same goes for the \lstinline{functools.partial} function, which allows getting function objects with partially assigned parameters.
The Pickle scanners are configured to flag these functions as unsafe; this translates to a high volume of files wrongly flagged as unsafe on Hugging Face (about 98\%!).
MalHug \cite{malhug} also identified this issue, and they manually confirmed that none of these flagged files were actually attempting to cause RCE.

\section{Threat Model}
\label{sec:threat_model}
Our threat model is based on realistic threat scenarios discussed previously in security articles and research papers: model serialization attacks that utilize the PDRE \cite{AWS_AISEC_PICKLE,malhug}.

\noindent \textbf{Model Deserialization Attacks:}
\label{sec:msa}
An attacker prepares an AI model that executes some malicious behavior upon deserialization. This can be achieved in PyTorch models based on the well-known Pickle exploit \cite{malhug} and also in TensorFlow models \cite{zhu2024my}. The attacker can then directly distribute the malicious model via model hubs such as the widely-used HuggingFace \cite{Huggingfacecite} for unsuspecting users to download and execute.
Victim users can easily download and run these models, and we argue that the majority of AI model users don't thoroughly inspect models before executing them, resulting in RCE on the machine of the user.
Past articles show that attackers can potentially upload malicious content to HuggingFace under the guise of trusted vendors using leaked API keys \cite{hf_token_leak}, or by cleverly hijacking existing trusted and widely-used repositories in a manner similar to phishing attacks \cite{hf_aijacking}. Hence, merely mistrusting unknown publishers might not be sufficient to ensure safety.
Malicious models can also attack users indirectly by being a third-party dependency used by other software, similarly to software supply-chain attacks \cite{ohm2020backstabber}.
In this scenario, the user will use a repository that in itself doesn't pose a danger, but it depends on another repository that does.

\section{Methodology}
\label{methodology}
\begin{figure}
    \centering
    \includegraphics[width=1\linewidth]{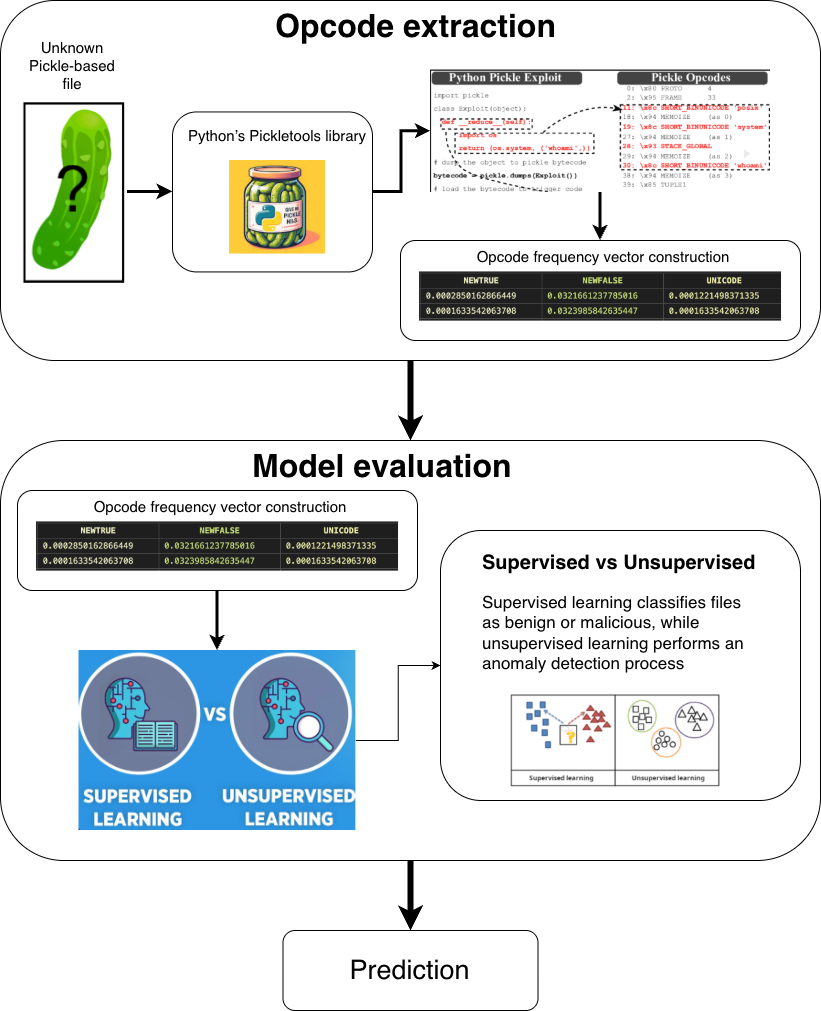}
    \caption{Overview of the proposed detection pipeline. In the first stage, the Pickle-based model file is parsed using Python’s Pickletools library, and the extracted opcodes are converted into normalized opcode-frequency vectors. In the second stage, these vectors are evaluated using supervised or unsupervised learning techniques to determine whether the file is benign or malicious, ultimately producing a final prediction}
    \label{fig:safepickle_methodology}
\end{figure}

Our objective is to identify malicious payloads hidden within Pickle-based machine learning models without executing them. To achieve this, we extract opcodes from the serialized files using Python’s Pickletools library \cite{pickletools} and transform them into structured feature representation that capture the internal behavior of the Pickle-based file deserialization process. As shown in the process pipeline in Figure~\ref{fig:safepickle_methodology}, the overall workflow is divided into three stages: data collection, feature extraction, and model training.

\subsection{Feature Extraction}
\label{Feature Extraction}
Opcode sequences were extracted using Python’s \texttt{pickletools} module \cite{pickletools}, which provides a disassembled representation of the serialized instructions. Each Pickle-based file was represented using an opcode distribution vector.

\paragraph{\textbf{Opcode Frequencies}}  
Each of the vector's dimension is an opcode appearing percentage calculated using the formula below:

\[
f_i = \frac{\text{count}(\text{opcode}_i)}{\sum_j \text{count}(\text{opcode}_j)}, 
\qquad \sum_i f_i = 1
\]
where \(i\) denotes the opcode of interest and \(j\) ranges over all opcodes present in the file.  
The denominator aggregates the total number of opcode occurrences, ensuring that the resulting vector is normalized and invariant to file size.  
This representation captures the structural behavior of the Pickle program independently of the model’s actual content.

\begin{figure}[H]
    \centering
    \caption{t-SNE representation of our curated dataset}
    \includegraphics[width=\linewidth]{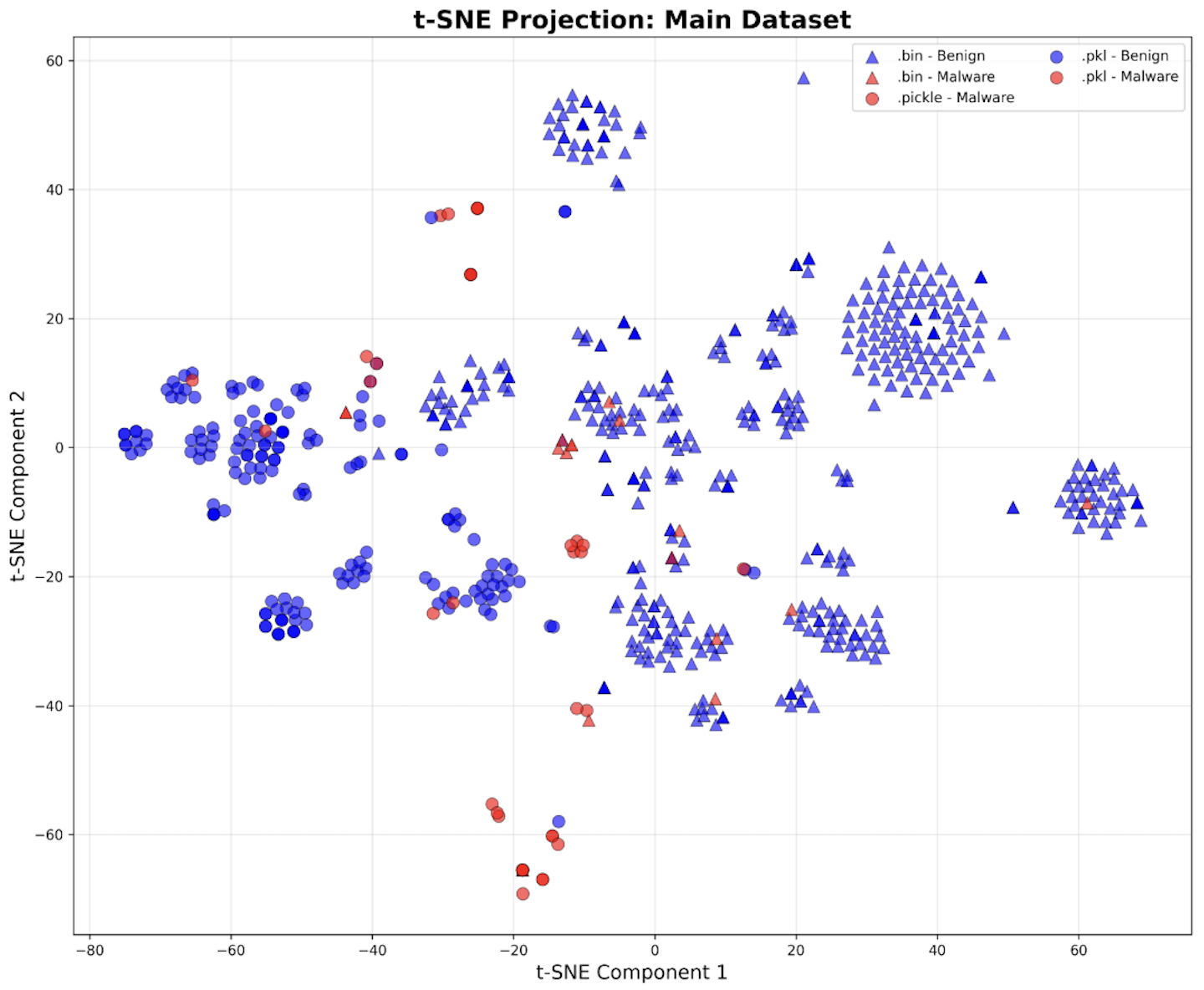}
    \label{fig:tsne_curated}
\end{figure}

\begin{figure}[H]
    \centering
    \caption{t-SNE representation of the PickleBall dataset}
    \includegraphics[width=\linewidth]{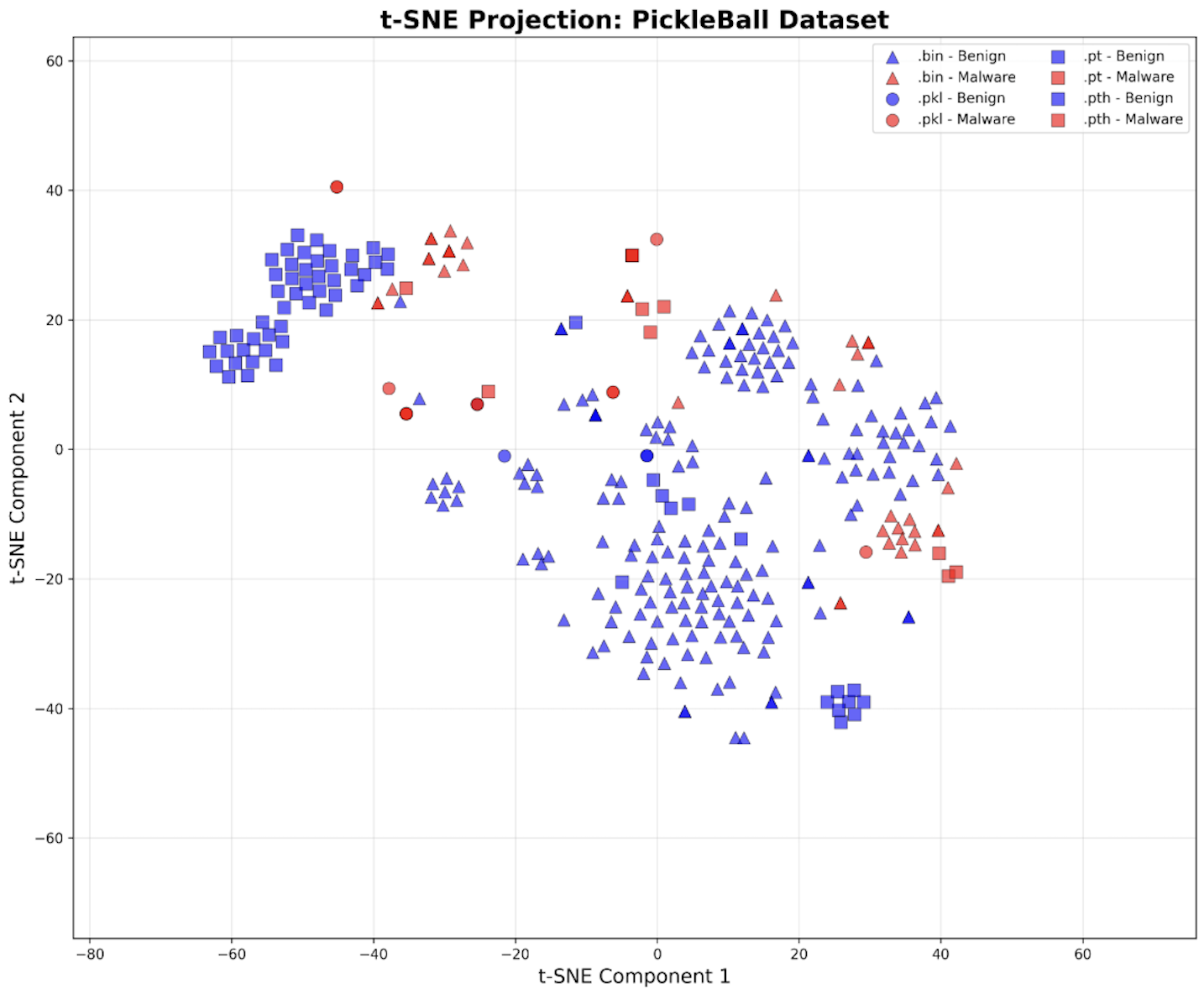}
    \label{fig:tsne_pb}
\end{figure}

\paragraph{\textbf{Feature Representation}}
In this study, we exclusively use the opcode distribution vectors as the feature representation.  
By relying solely on normalized opcode frequencies, the framework provides a direct, interpretable description of the internal Pickle structure.  
This enables the detection of structural anomalies characteristic of malicious behavior without requiring embeddings, handcrafted statistical descriptors, or additional engineered features.
Opcode dimensions that were unused across the entire dataset (all-zero columns) were removed to eliminate noise and redundant features. In addition, we represented our curated dataset and PickleBall dataset  using t-SNE projections which is a nonlinear dimensionality-reduction method for visualizing high-dimensional data in 2D. As illustrated in Figure~\ref{fig:tsne_curated} and Figure~\ref{fig:tsne_pb}, the curated dataset exhibits a markedly broader dispersion within the embedding space, indicative of increased structural heterogeneity. In contrast, the PickleBall dataset demonstrates more compact clustering patterns, reflecting comparatively reduced variability.

\subsection{Model Training}
The purpose of the model training phase is to learn structural patterns in Pickle-based files that distinguish benign models from malicious ones.  
All models were trained exclusively on the opcode distribution vectors extracted during static analysis, following the feature extraction described in Section~\ref{Feature Extraction}.  
We employed both supervised classifiers and unsupervised anomaly detectors to support detection in labeled and zero-day scenarios.

To detect malicious Pickle-based models, we employed both supervised and unsupervised learning, each using a distinct set of models.

\noindent \textbf{Supervised learning}: we trained three classifiers-\texttt{RandomForest}, \texttt{CatBoost}, and an \texttt{Autoencoder}-based classifier on labeled opcode distribution vectors.  
These models learned discriminative decision boundaries that separate benign from malicious files based on structural patterns in their opcode distributions.

\noindent \textbf{Unsupervised learning}: we applied anomaly detection methods including \texttt{LocalOutlierFactor}, \texttt{IsolationForest}, and \texttt{OneClassSVM}.  
These models were trained exclusively on benign opcode distributions and flagged samples as malicious when they deviated significantly from normal structural patterns.

\subsection{Security Tools}
In the following subsections, we introduce security tools that we used during our experimentation.
\subsubsection{Pickle2py: Converting Pickles to Python Code}
\label{sec:pickle2py}

We design a method for converting Pickles to Python code. This transformation from Pickle, a specialized format, to Python - a well-known programming language with lots of parsing/analysis libraries, etc., greatly enhances our ability to analyze the Pickles.
We call this method \textbf{Pickle2py}.
This idea also appears in MalHug \cite{malhug}, however, they do not provide the code that does this.
The Pickle2py tool enables deep inspection of Pickle file content, as opposed to the past methodology of Pickle API scanning (see Section \ref{sec:pickle_api_scanners}).  

To implement the Pickle2py functionality, we utilize existing open-source Python libraries.
First, we repurpose the Picklescan code that "walks" a file and iteratively locates Pickle files within it.
Then, we use the Fickling \cite{Fickling} library to construct an Abstract Syntax Tree (AST) from the raw Pickle bytes, and finally, we use the Python builtin \lstinline{ast.unparse} function to produce a string that contains the Python program that equivalently does what unpickling the Pickle would've executed. This has the significant advantage that it does not unpickle the files, hence, it has no danger of RCE.
The Pickle2py method converts the problem of analyzing a Pickle file (custom format) to analyzing Python source code, which is much better covered. For example, LLMs usually can parse Python code, so this allows analyzing the Pickles using LLMs.

\section{Evaluation}
\label{evaluation}
We evaluated our detection framework to determine its effectiveness, generalization ability, and practical suitability for identifying malicious Pickle-based files. This section outlines the metrics used and the experimental setups designed to assess performance across diverse datasets and threat scenarios.

In our experiments, we used the Scikit-Learn \cite{scikit-learn} library version 1.7.2 for all the ML preprocessing, training, and inference.
We used the default hyperparameters in all our experiments in our classifiers, we leave the optimization of hyperparameters to future work.

\subsection{Experimental Setup}
All experiments are run on a Ubuntu Linux 24.04 server, equipped with an Intel(R) Xeon(R) w5-2445 CPU (20 cores @ 3.10GHz), and 128 GB RAM.


\subsection{Data Collection}
Data collection posed significant challenges due to the limited availability of reliably labeled datasets. Labeling Pickle-based files typically requires the use of existing analysis tools, but however, none of these existing tools can be considered fully trustworthy. Consequently, relying on imperfect labels introduces the risk of contaminating the dataset with inaccuracies, which in turn may negatively impact model performance. To address this limitation, we utilized four datasets-summarized in Table~\ref{tab:dataset_summary}-comprising Pickle-based machine learning files selected to capture a broad spectrum of benign behaviors, adversarial constructs, and malicious samples. These datasets were either manually labeled by us or sourced from prior peer-reviewed studies in which ground-truth labels were provided.

\paragraph{\textbf{Our Curated Dataset from Hugging Face}}  
The first and main dataset consists of 648 benign and 79 malicious models we collected from the Hugging Face model hub, spanning diverse application domains such as computer vision, natural language processing, and decision support systems. In order to label them, all files were inspected using our static analysis tool \texttt{Pickle2py} (see Section \ref{sec:pickle2py}), which reconstructs the Python code embedded within the Pickle serialization.  
This enables the identification of operations that are unnecessary for model deserialization and the detection of potential malicious constructs.  
By relying solely on static reconstruction and avoiding execution, \texttt{Pickle2py} ensures safe inspection of both benign and malicious files and provides a reliable foundation for subsequent opcode-based analysis.

\paragraph{\textbf{PickleBall Dataset}}  
PickleBall~\citep{PickleBall2025} framework generates safe deserialization policies through static library analysis, making this dataset particularly suitable for benchmarking our learning-based approach and evaluating cross-dataset generalization. In their experiments they used their own curated dataset which contains 255 benign and 82 malicious Pickle-based machine learning files.

\paragraph{\textbf{Hide-and-Seek Dataset}}  
Hide-and-Seek paper~\citep{hide_and_seek} present techniques to create files that SOTA scanner fails to scan so we manually recreated those files and tested them using our methods. Each sample replicates or extends the deserialization attack mechanisms described in the study, including polyglot and multi-format payloads (e.g. \texttt{zip→pkl}, \texttt{bz2→pkl}, \texttt{tar→pkl}, \texttt{xz→pkl} and more) designed to evade traditional static scanners.  
These controlled constructions serve as research-grade adversarial test cases for evaluating zero-day robustness and scanner resilience to novel, unseen Pickle-based attack vectors.

\paragraph{\textbf{Malware Dataset}}  
We used also six real malicious model files containing executable payloads embedded directly within Pickle-based serialization formats. Four of these samples originate from a publicly documented incident reported by ReversingLabs~\cite{reversinglabs}, which identified malicious uploads to Hugging Face involving both Pickle and PyTorch variants of two compromised model repositories. These models contained early-trigger reducer functions and corrupted Pickle streams specifically crafted to evade static scanning while still enabling code execution during loading.
The remaining sample is a file found on Hugging Face, it is a \href{https://huggingface.co/aisecre/EXP/blob/main/joblib/custom_undetected_model.joblib}{malicious joblib pickle file} that is flagged suspicious by ClamAV, but all other scanners label as safe. Moreover, the pickle import list looks just like a typical import list for a scikit-learn ML model; however, when you load the model using \lstinline{joblib}, a \lstinline{os.system} command is executed.
We chose this sample because the Hugging Face scanners missed it.

\begin{table}[H]   
\centering
\begin{tabular}{lcc}
\hline
\textbf{Dataset} & \textbf{Benign} & \textbf{Malicious} \\ \hline
Curated (Hugging Face) & 648 & 79 \\
PickleBall \cite{PickleBall2025} & 255 & 82 \\ 
Hide and Seek \cite{hide_and_seek} & -- & 9 \\ 
Malware & -- & 5 \\ \hline
\end{tabular}
\caption{Overview of Pickle-based datasets used in this study. \\}
\label{tab:dataset_summary}
\end{table}

\subsection{Metrics}
We evaluate our models using true positives (TP), true negatives (TN), and the F1-score. These metrics were selected because they directly reflect the security requirements of Pickle-based malware detection. TP and TN allow us to measure how well the model distinguishes malicious from benign files, which is essential for assessing both detection capability and normal-operation behavior. Since the dataset may be imbalanced, accuracy alone would be misleading. Therefore, we use the F1-score as a balanced measure that captures the trade-off between precision and recall. This ensures that both missed malicious samples and unnecessary alerts are properly accounted for during evaluation.

\subsection{Evaluation Strategies}
To comprehensively assess the effectiveness and robustness of our framework, we conducted multiple evaluation experiments encompassing both supervised and unsupervised learning paradigms, as well as cross-dataset and runtime analyses: \\

\begin{enumerate}[
    label=\textbf{Exp\arabic*.},
    leftmargin=0pt,
    labelsep=0.5em,
    itemindent=2em,
    listparindent=0pt,
    align=left,
    nosep,
    itemsep=1.0\baselineskip, 
    before=\raggedright
]

    \item \textbf{Evaluation on our dataset.}  
    In this experiment, our models were trained and tested on our curated dataset to establish baseline performance and validate their ability to detect malicious Pickle-based files. In parallel, we compared our results against existing malware detection tools such as \texttt{ModelScan}, \texttt{Fickling}, and other open-source scanners to benchmark detection accuracy and reliability.

    \item \textbf{Evaluation on the PickleBall Dataset.} 
    The same experimental procedure was applied to the PickleBall dataset. We further compared our framework’s results with tools used in the PickleBall paper, including the \texttt{PickleBall} scanner itself, to assess consistency and performance across independently sourced data.

    \item \textbf{Cross-Dataset Evaluation.}  
    Supervised models trained on our curated dataset were tested on external datasets to evaluate generalization and zero-day detection capability. This evaluation consisted of three setups:  
    (1) training on our curated dataset and testing on the PickleBall dataset,  
    (2) training on our curated dataset and testing on the Hide-and-Seek dataset, and  
    (3) training on our curated dataset and testing on the malware dataset containing stealthy and scanner-evasive samples.

    \item \textbf{Runtime Analysis.}  
    To assess deployment practicality, we measured the average runtime per file for our framework and compared it with existing scanners, including \texttt{PickleBall} scanner, and other open-source model security tools.  
    Our runtime measurements encompass the complete detection pipeline, including opcode extraction, feature generation, and model prediction, thereby ensuring a fair and consistent comparison of computational efficiency across all evaluated methods.  
    This analysis underscores the balance between runtime performance and detection recall and F1-score achieved by our framework in relation to existing scanners. 
\end{enumerate}

\subsubsection{Baseline scanners}
To contextualize the performance of our approach, we evaluate a set of widely used security tools as baseline scanners that operate directly on the raw serialized pickle-file dataset. Importantly, Pickleball's own scanning mechanism is \emph{not} used as a baseline here. We only use it within
the Pickleball specific experiments following the procedure described in the original repository.  
For all other experiments, the following baseline tools were employed with fixed versions to ensure reproducibility:

\begin{itemize}

    \item \textbf{picklescan} (v0.0.30) \cite{picklescan}: 
    A compact static inspection tool tailored for Python pickle files. 
    It highlights suspicious opcode usage and identifies serialization patterns that frequently appear in exploit-bearing pickles. 
    In our evaluation, the tool's "Infected files" field is interpreted as a binary malicious/benign signal.

    \item \textbf{ClamAV} (v1.5.1, built from the Cisco, Talos repository at commit \texttt{c73755d}\cite{clamav}): 
    The experiment utilizes a source-compiled installation of ClamAV, rather than distribution-packaged or Python-wrapped variants. 
    As a signature-driven antivirus engine designed for traditional malware detection, ClamAV does not directly target pickle-specific threats; however, it provides an informative baseline for assessing the generalizability of conventional malware scanners when confronted with serialized Python artifacts.

    \item \textbf{Fickling} (Git commit \texttt{3a46746}\cite{model-fickling}): 
    A security-oriented pickle decompiler and static analysis framework. 
    Fickling reconstructs the internal structure of pickle bytecode and surfaces potentially hazardous constructs, including dynamic execution pathways (\texttt{eval}, \texttt{exec}) and invocation of system-level functions such as \texttt{os.system} or \texttt{subprocess}. 
    Its rule-based warnings serve as indicators of unsafe or malicious deserialization behavior.

    \item \textbf{ProtectAI ModelScan} (Git commit \texttt{abc4b15}\cite{modelscan}): 
    A model-focused security scanner built to examine machine-learning serialization formats. 
    The specific version used corresponds to commit \texttt{abc4b15} from the official GitHub repository,
    Any non-zero value in the tool's "Total Issues" field is treated as an affirmative detection of unsafe content within the model artifact. 
    
    

    \item \textbf{VirusTotal} (v1.1.0, accessed via the \texttt{vt-py} Python API): 
    For each file, an MD5 hash is computed and used to query VirusTotal's aggregated multi-engine threat-intelligence service. 
    This hash-only lookup avoids uploading the full artifact, instead returning any previously recorded detections from the dozens of antivirus engines integrated into the platform. 
    A sample is labeled malicious if at least one backend engine has historically flagged the corresponding MD5 hash.
   
\end{itemize}

Each scanner is executed independently on every file in the dataset. For every baseline tool we record both the binary detection output (malicious, benign or failed) and the wall-clock run time per file. All results are aggregated and exported for further comparison against our method.

\subsection{Results}   
\label{sec:results}
This section presents a comprehensive evaluation of the proposed framework across several experimental settings, including supervised and unsupervised learning on individual datasets, cross-dataset generalization, comparison with state-of-the-art malware detection tools, and an analysis of runtime efficiency. All experiments were conducted using the baseline opcode-distribution vectors described in Section~\ref{Feature Extraction}.

Performance is evaluated using three primary metrics: True Positive (TP) Rate, True Negative (TN) Rate, and F1-score. For clarity, within each experimental category (Supervised, Unsupervised, and SOTA Scanners), the highest value for each metric is highlighted in \textbf{bold}. In addition, any F1-score that exceeds the best F1-score reported by the state-of-the-art scanners is marked with \textbf{\underline{bold underline}}, signifying that the proposed framework achieves performance beyond the current benchmark. \\

\begin{enumerate}[
    label=\textbf{Exp\arabic*.},
    leftmargin=0pt,
    labelsep=0.5em,
    itemindent=2em,
    listparindent=0pt,
    align=left,
    nosep,
    itemsep=1.0\baselineskip, 
    before=\raggedright
]

    \item \textbf{Evaluation on the Curated Dataset.}  
    We first evaluated both supervised and unsupervised models on our curated dataset using standard train-test splits. This setup provides a controlled environment for measuring baseline detection performance before introducing external variability.

    \begin{table}[H]
    \centering
    \caption{\textsc{Comparison of SafePickle Models and SOTA Scanners on our Curated Dataset.}}
    \label{tab:curated_comparison}
    \renewcommand{\arraystretch}{1.2}
        
    \resizebox{\columnwidth}{!}{%
    \begin{tabular}{lccc|ccc|ccccc}
    \toprule
    \textbf{Model} & \multicolumn{3}{c|}{\textbf{Unsupervised}} & \multicolumn{3}{c|}{\textbf{Supervised}} & \multicolumn{5}{c}
    {\textbf{SOTA Scanners}} \\
    \cmidrule(lr){2-4} \cmidrule(lr){5-7} \cmidrule(lr){8-12}
     & LOF & IF & OC-SVM & RF & AE & CB & PS & C-AV & FIC & MS & VT \\ 
    \midrule
    \rowcolor[HTML]{F8F8F8}
    \textbf{TP (\%)} 
        & 87.50 & \textbf{89.50} & 87.50 
        & \textbf{87.50} & 81.25 & \textbf{87.50} 
        & 97.47 & 3.80 & 57.14 & \textbf{100.00} & 7.69 \\
        
    \textbf{TN (\%)} 
        & \textbf{90.77} & 87.69 & 86.15 
        & \textbf{96.92} & 94.62 & 93.08 
        & 0.00 & \textbf{99.85} & 85.71 & 0.00 & 99.77 \\
        
    \rowcolor[HTML]{F8F8F8}
    \textbf{F1 (\%)} 
        & \underline{\textbf{66.67}} & 60.87 & 58.33 
        & \underline{\textbf{90.01}} & 86.68 & 83.72 
        & 19.15 & 7.23 & \textbf{62.75} & 18.66 & 14.04 \\
    \bottomrule
    \end{tabular}%
    }
    \end{table}
    
    As shown in Table~\ref{tab:curated_comparison}, SafePickle’s supervised models-most notably \texttt{RandomForest}-deliver the strongest overall performance, achieving the best balance of true positives, true negatives, and F1-score. This outcome is expected, as supervised learning benefits from direct exposure to labeled benign and malicious samples, enabling the models to learn precise decision boundaries.
    In contrast, the unsupervised methods naturally exhibit lower absolute performance. However, \texttt{LocalOutlierFactor} still stands out as the most effective anomaly-based detector, confirming its suitability for identifying deviations without prior class labels. Traditional scanners show a markedly inconsistent pattern: tools such as \texttt{PickleScan} and \texttt{ModelScan} detect many malicious files but suffer from high false-positive rates, whereas \texttt{ClamAV} and \texttt{VirusTotal} miss a substantial portion of threats. The comparatively higher and more stable F1-scores of the SafePickle models underscore their robustness and reliability across diverse threat conditions.

    \item \textbf{Evaluation on the PickleBall Dataset.} 
    Our supervised and unsupervised models were also trained on the PickleBall dataset. As demonstrated in Figure~\ref{fig:tsne_pb}, the PickleBall dataset exhibits substantially lower diversity. Consequently, we expect the strongest SafePickle models, along with the SOTA scanners, to achieve improved performance under these conditions.

    \begin{table}[H]
    \centering
    \caption{\textsc{Comparison of SafePickle Models and SOTA Scanners on PickleBall Dataset.}}
    \label{tab:pickleball_comparison}
    \renewcommand{\arraystretch}{1.2}

    \resizebox{\linewidth}{!}{
    \begin{tabular}{lccc|ccc|cccccc}
    \toprule
    \textbf{Model} 
        & \multicolumn{3}{c|}{\textbf{Unsupervised}} 
        & \multicolumn{3}{c|}{\textbf{Supervised}} 
        & \multicolumn{6}{c}{\textbf{External Scanners}} \\
        
    \cmidrule(lr){2-4}
    \cmidrule(lr){5-7}
    \cmidrule(lr){8-13}

     & LOF & IF & OC-SVM & RF & AE & CB & PS & C-AV & FIC & MS & VT & PB \\ 
    \midrule
    
    \rowcolor[HTML]{F8F8F8}
    \textbf{TP (\%)} 
        & 42.86 & \textbf{71.43} & \textbf{71.43}
        & 85.90 & 85.71 & \textbf{92.86}
        & 97.18 & 26.76 & 88.24 & \textbf{100.00} & 7.69 & \textbf{100.00} \\
        
    \textbf{TN (\%)} 
        & \textbf{80.77} & 78.85 & 73.08
        & \textbf{96.15} & 96.15 & 94.23
        & 92.55 & \textbf{99.15} & 57.14 & 0.00 & 92.77 & 79.80 \\
        
    \rowcolor[HTML]{F8F8F8}
    \textbf{F1 (\%)} 
        & 40.00 & \textbf{57.14} & 52.63
        & 90.94 & 90.93 & \underline{\textbf{91.37}}
        & \textbf{86.79} & 41.35 & 54.88 & 39.14 & 11.81 & 76.09 \\
    
    \bottomrule
    \end{tabular}
    }
    \end{table}
    
    As shown in Table~\ref{tab:pickleball_comparison}, the supervised SafePickle models, particularly \texttt{RandomForest} and \texttt{CatBoost}, provide strong and well-balanced performance while maintaining high true-positive, true-negative rates and F1-scores under distribution shift. This result is expected, as supervised approaches are trained directly on labeled benign and malicious samples. The unsupervised models, including \texttt{IsolationForest} and \texttt{OC-SVM}, achieve moderate yet meaningful performance. Even without access to labeled data, these models outperform several SOTA scanners, which highlights the value of anomaly-detection techniques.
    
    External scanners show a less stable pattern. Signature-based tools such as \texttt{PickleScan} and \texttt{ModelScan} correctly identify many malicious files but generate a large number of false positives. In contrast, traditional antivirus systems (for example, \texttt{ClamAV} and \texttt{VirusTotal}) achieve high true-negative rates but fail to detect a substantial portion of malicious samples. Overall, the consistently higher F1-scores of the SafePickle models point to more reliable performance across different threat types and emphasize their robustness when evaluated on independently sourced data.

    \item \textbf{Cross-Dataset Evaluation.}  
    To assess generalization and zero-day detection capability, supervised SafePickle models trained exclusively on our curated dataset were evaluated on three external datasets representing distinct and increasingly challenging threat types:

    \begin{enumerate}
        \item \textbf{PickleBall Dataset:}  
        The models were evaluated on the PickleBall dataset to verify that, although they were trained on our curated data, they are sufficiently generalizable to handle previously unseen samples. This evaluation assesses whether the models can maintain high true-positive and true-negative rates, as well as strong F1-scores, when applied to new and independent data.
    
        \item \textbf{Hide-and-Seek Dataset:}  
        SafePickle supervised models were also evaluated on nine adversarial samples reconstructed from the \texttt{Hide-and-Seek} paper. These samples include polyglot, multi-format, and compressed variants that were deliberately designed to trigger failures in SOTA scanners.

        \item \textbf{Malware Dataset:}  
        Models were further evaluated on a set of six operational malicious Pickle files containing fully functional attack payloads. Given the small size and heterogeneous nature of this dataset, we report absolute detection counts rather than percentages.
        
    \end{enumerate} 
    
    Together, these datasets provide a rigorous evaluation of SafePickle's ability to generalize across distribution shifts, adversarial transformations, and malicious payloads.\\

    As shown in Table~\ref{tab:cross_pickleball}, the supervised SafePickle models, particularly \texttt{RandomForest}, achieve strong and stable detection performance under distribution shift. The \texttt{Autoencoder} obtains the highest true-negative rate among the SafePickle models, indicating effective discrimination of benign samples in previously unseen data. External scanners exhibit polarized behavior, with some identifying nearly all malicious files while performing poorly on benign detection, and others missing the majority of threats altogether.
    \begin{table}[H]
    \centering
    \caption{\textsc{Cross-Dataset Evaluation on the PickleBall Dataset}}
    \label{tab:cross_pickleball}
    \renewcommand{\arraystretch}{1.2}
    \setlength{\tabcolsep}{6pt}
    
    \resizebox{\linewidth}{!}{
    \begin{tabular}{lccc|cccccc}
    \toprule
        \textbf{Model} 
        & \multicolumn{3}{c|}{\textbf{SafePickle Supervised}} 
        & \multicolumn{6}{c}{\textbf{SOTA Scanners}} \\
        
    \cmidrule(lr){2-4}
    \cmidrule(lr){5-10}
    
        & RF & AE & CB 
        & PS & C-AV & FIC & MS & VT & PB \\
    \midrule
    
    \rowcolor[HTML]{F8F8F8}
    \textbf{TP (\%)} 
        & \textbf{93.0} & 57.9 & \textbf{93.0}
        & 97.18 & 26.76 & 88.24 & \textbf{100.00} & 7.69 & \textbf{100.00} \\
    
    \textbf{TN (\%)} 
        & 60.8 & \textbf{87.84} & 60.4
        & 92.55 
        & \textbf{99.15} & 57.14 & 0.00 & 92.77 & 79.80 \\
    
    \rowcolor[HTML]{F8F8F8}
    \textbf{F1 (\%)} 
        & \textbf{81.22} & 75.12 & 81.15
        & \underline{\textbf{86.79}} 
        & 41.35 & 54.88 & 39.14 & 11.81 & \textbf{76.09} \\
    
    \bottomrule
    \end{tabular}
    }
    \end{table}

    \begin{table*}[t]
    \centering
    \caption{Scope of Pickle-Based Malware Detection Across File Loading Paths \\ 
    (Notation: \CIRCLE = detected, \LEFTcircle = scanned-not detected, \Circle = failed scan)}
    \label{tab:loading_paths}
    \setlength{\tabcolsep}{6pt}
    
    \resizebox{\textwidth}{!}{%
    \begin{tabular}{l|ccc|ccccc}
    \hline
    \textbf{Path} &
    \textbf{RF} &
    \textbf{CatBoost} &
    \textbf{AE} &
    \textbf{PickleScan} &
    \textbf{ClamAV} &
    \textbf{Fickling} &
    \textbf{ModelScan} &
    \textbf{VirusTotal} \\ 
    \hline
    
    pkl & \CIRCLE & \CIRCLE & \CIRCLE & \CIRCLE & \CIRCLE & \CIRCLE & \CIRCLE & \CIRCLE \\
    
    zip $\rightarrow$ pkl 
        & \CIRCLE & \CIRCLE & \CIRCLE & \CIRCLE & \Circle & \Circle & \CIRCLE & \Circle \\
    
    zip $\rightarrow$ zip $\rightarrow$ pkl 
        & \CIRCLE & \CIRCLE & \LEFTcircle & \Circle & \Circle & \Circle & \Circle & \Circle \\
    
    tar $\rightarrow$ pkl 
        & \CIRCLE & \LEFTcircle & \LEFTcircle & \Circle & \Circle & \Circle & \Circle & \LEFTcircle \\
    
    bz2 $\rightarrow$ pkl 
        & \CIRCLE & \CIRCLE & \LEFTcircle & \Circle & \Circle & \Circle & \Circle & \Circle \\
    
    gz $\rightarrow$ pkl 
        & \CIRCLE & \CIRCLE & \LEFTcircle & \Circle & \Circle & \Circle & \Circle & \Circle \\
    
    zlib $\rightarrow$ pkl 
        & \CIRCLE & \CIRCLE & \LEFTcircle & \Circle & \Circle & \Circle & \Circle & \Circle \\
    
    lz4 $\rightarrow$ pkl 
        & \CIRCLE & \CIRCLE & \LEFTcircle & \Circle & \Circle & \Circle & \Circle & \Circle \\
    
    lzma $\rightarrow$ pkl 
        & \CIRCLE & \CIRCLE & \LEFTcircle & \Circle & \Circle & \Circle & \Circle & \Circle \\
    
    xz $\rightarrow$ pkl 
        & \CIRCLE & \CIRCLE & \LEFTcircle & \Circle & \Circle & \Circle & \Circle & \Circle \\
    
    \hline
    \end{tabular}
    }
    \end{table*}
    
    Table~\ref{tab:loading_paths} evaluates detection robustness across compressed, wrapped, and multi-stage file-loading paths. SafePickle’s supervised models detect every malicious sample across nearly all transformations, including deeply nested and compressed formats. Traditional scanners succeed almost exclusively on plain \texttt{pkl} files and fail consistently on alternative encodings, showing limited resilience to wrapper-based evasion. \\
    
    Finally, models were evaluated on five operational malicious Pickle files.
    All three SafePickle models-\texttt{RandomForest}, \texttt{Autoencoder}, and \texttt{CatBoost}-successfully detected all five threats.  
    Among traditional scanners, \texttt{PickleScan} and \texttt{ModelScan} detected 4/5.  
    \texttt{Fickling} identified only 1/5 samples, and both \texttt{ClamAV} and \texttt{VirusTotal} detected none.
    
    These results demonstrate that SafePickle reliably identifies malicious Pickle payloads, whereas existing scanners exhibit substantial blind spots and fail to generalize beyond narrow, signature-matched threats.

    \item \textbf{Runtime Analysis.}  
    We evaluated runtime performance to assess the feasibility of deploying SafePickle in real-world scanning pipelines.  
    Measurements include opcode extraction, feature construction, and model inference, providing a fair comparison against existing scanners.

    \begin{table}[H]
    \centering
    \caption{\textsc{Average Runtime per File for Detection Methods}}
    \label{tab:runtime}
    \renewcommand{\arraystretch}{1.2}
    
    \begin{tabular}{l c}
    \toprule
        \textbf{Scanner / Model} & \textbf{Avg. Runtime (ms)} \\
    \midrule
    
        \multicolumn{2}{c}{\textbf{SOTA Scanners}} \\
    \midrule
    
    \rowcolor[HTML]{F8F8F8}
        PickleScan   & 167.8 \\
        ClamAV     & 11,875.6 \\
    \rowcolor[HTML]{F8F8F8}
        Fickling     & \textbf{44.4} \\
        ModelScan    & 2,776.8 \\
    \rowcolor[HTML]{F8F8F8}
        VirusTotal   & 3,081.9 \\
        PickleBall   & 14,000.42 \\
        
    \midrule
        \multicolumn{2}{c}{\textbf{SafePickle}} \\
    \midrule
    
    \rowcolor[HTML]{F8F8F8}
        \textit{\textbf{RandomForest}} & 0.912 \\
        \textit{\textbf{CatBoost}}     & \textbf{0.016} \\
    \rowcolor[HTML]{F8F8F8}
        \textit{\textbf{Autoencoder}}  & 0.019 \\
    
    \bottomrule
    \end{tabular}
    \end{table}

    Traditional scanners incur substantial processing overhead, often requiring hundreds to thousands of milliseconds per file as shown in Table~\ref{tab:runtime}, due to heavyweight signature matching, archive unpacking, or cloud-assisted analysis. Even the fastest legacy tool, \texttt{Fickling}, remains significantly slower than any SafePickle model.

    SafePickle delivers consistently low inference latency, with all models operating well below one millisecond per file. \texttt{CatBoost} achieves the fastest runtime (0.016 ms), while both \texttt{RandomForest} and the \texttt{Autoencoder} also maintain sub-millisecond performance. This makes the framework suitable not only for offline batch scanning but also for high-throughput environments and latency-sensitive workflows such as CI/CD pipelines, cloud inference gateways, or user-facing applications.

    Overall, SafePickle combines strong detection performance with dramatically reduced computational cost, offering a practical and scalable alternative to traditional scanning tools.
\end{enumerate}

\subsection{Summary of Observations}
Across all evaluation scenarios, including curated datasets, cross-dataset generalization, adversarial polyglot inputs, real-world malicious Pickle-based files, and runtime analysis, \textbf{SafePickle consistently outperforms all existing SOTA scanners}.

The supervised models, with \texttt{RandomForest} as the strongest performer, show stable and reliable behavior across every dataset tested. They maintain high true-positive and true-negative rates under substantial distribution shifts and achieve strong F1-scores. This balance between precision and recall is particularly important in security settings where class imbalance often makes accuracy an unreliable metric.

Traditional scanners, in comparison, exhibit inconsistent and polarized behavior. Some detect most malicious files but misclassify many benign ones, while others overlook the majority of threats. SafePickle does not suffer from these limitations and remains effective even on adversarial, obfuscated, compressed, or nested malicious inputs. Its opcode-based representation enables the system to capture behavioral patterns that remain consistent across different file formats and wrapping methods.

The models also generalize well to operational malicious Pickle-based files. Every supervised SafePickle model successfully detected all samples in this evaluation, whereas traditional scanners recognized only a portion of them. Alongside its strong detection metrics, SafePickle provides exceptional runtime efficiency.
It operates in the sub-millisecond range, making it several orders of magnitude faster than existing SOTA scanners. This level of efficiency is crucial for practical deployments that require rapid scanning or integration into high-throughput systems. The system further maintains robustness across different serialization paths and file-wrapping strategies.
Overall, the results demonstrate that SafePickle delivers a \textbf{highly accurate, generalizable, and computationally efficient} defense mechanism for Pickle-based malware detection. It offers clear advantages over signature-based and static analysis tools.

\subsection{Security Limitations of PickleBall}
PickleBall provides a policy-based defense against malicious Pickle deserialization by statically analyzing machine-learning library source code to generate safe deserialization policies. While the approach offers an improvement over unguarded loading, it suffers from several inherent security and operational limitations that weaken its effectiveness in practice.

A core issue lies in PickleBall’s implicit trust in the library source code used to generate its policies. If a dependency is compromised-through supply-chain attacks, dependency poisoning, or malicious contributors-the policy generator will treat attacker-controlled callables as legitimate. This “trusted but poisoned” scenario silently subverts the entire defense, allowing arbitrary code execution during model loading while appearing compliant with the generated policy.

Beyond trust concerns, the static analysis itself provides incomplete coverage. Python’s highly dynamic behavior prevents PickleBall from identifying many legitimate callables, causing a significant portion of benign models to fail loading under the generated policies. Such under-approximation forces users to manually relax policies or whitelist additional callables, reducing the intended security guarantees. Moreover, PickleBall’s evaluation relies on the assumption that popular publicly available models are benign-a fragile assumption that risks overlooking already-infected models in the wild.

Operationally, PickleBall introduces substantial maintenance overhead. Policies must be regenerated for every library update, and the analysis requires non-trivial tooling such as Joern, Dockerized environments, and extensive source-tree processing. This reduces deployability in production pipelines and increases the likelihood of stale or inconsistent policies across environments.

Finally, even an ideal policy cannot eliminate all risks. Because PickleBall permits arbitrary compositions of “safe” operations, it remains vulnerable to logic bombs, behavioral triggers, or code-reuse attacks constructed from allowed callables. This residual attack surface may create a false sense of confidence while still enabling sophisticated exploits.

In summary, while PickleBall reduces some classes of risk, it depends on unverifiable trust assumptions, suffers from incomplete static coverage, and introduces notable operational complexity. It is best viewed as a partial safeguard within a broader defense-in-depth strategy that includes sandboxing, runtime monitoring, and safer serialization formats such as \texttt{SafeTensors}.

\section{Discussion and Future Works}
The experimental results collectively highlight both the limitations of existing Pickle security tools and the strengths of a learning-based, opcode-driven approach. Traditional scanners rely primarily on file signatures, shallow heuristics, or direct inspection of model structure. As demonstrated across multiple evaluations, these methods consistently fail when the Pickle payload is wrapped, compressed, archived, or embedded within multi-stage polyglot formats. SafePickle, by contrast, remains effective across such transformations because it learns behavioral patterns reflected in opcode distributions rather than depending on surface-level file characteristics.

Several directions emerge for further advancement of the SafePickle framework. Although the current feature engineering pipeline already leverages opcode-based representations with strong results, exploring richer structures-such as sequence-aware models or graph-based control-flow abstractions may further enhance expressiveness and allow the system to capture more nuanced malicious patterns.

Expanding SafePickle to additional serialization ecosystems represents another promising opportunity. Extending support to formats such as \texttt{HDF5}, \texttt{Keras}, \texttt{ONNX}, or proprietary machine-learning serialization standards would increase its applicability in modern model-deployment pipelines. Evaluating the framework’s resilience to adversarially manipulated opcode distributions is also important for ensuring long-term robustness.

Incorporating SafePickle into a continuous learning pipeline may offer additional benefits by enabling periodic updates as new malware families emerge. Finally, hybrid detection strategies-combining SafePickle with lightweight rule-based or signature-based scanners-could unify behavioral and heuristic signals, potentially yielding stronger defenses than either approach alone.

Overall, these directions highlight natural next steps for improving the robustness, interpretability, and practical deployment readiness of SafePickle as a general-purpose defense mechanism for Pickle-based malware detection.

\section{Related Work}
\noindent \textbf{AI Model Attacks:} Previous research on attacks that use the AI model as a utility for attacking mainly focus on AI model steganography attacks, and AI model deserialization attacks. 

AI model steganography involves data hiding techniques, with works that develop novel attack patterns such as EvilModel \cite{wang2021evilmodel,wang2022evilmodel}, and MaleficNet \cite{maleficnet1,maleficnet2}. These works explore ways to craft malicious AI models with embedded malicious payloads that automatically extract and execute using some trigger mechanism, or by exploiting model deserialization RCE.
In response to the growing trend of AI model steganography attack research, defensive techniques that involve detection \cite{dl_steganalysis_1,danigil_stega_1} and zero-trust CDR \cite{gilkarov2025neupermdisruptingmalwarehidden} have emerged.

AI model deserialization attacks have been extensively researched, mostly in Pickle-based files. 
Marco Slaviero \cite{Pickle_serialization_vul} initially demonstrated in 2011 that an attacker can abuse Pickle serialization to execute arbitrary code or system operations - this sparked further research into attacks and defenses involving the PDRE.
To combat the PDRE, the restricting Unpickler was introduced \cite{PickleRestriction}, but it was quickly shown by works such as PainPickle \cite{huang2022pain} that the policy-based restricting Unpickler can be bypassed - necessitating more robust defenses.
Various Pickle-based file scanners or AI/ML model scanners were introduced, including Picklescan \cite{picklescan}, which detects possibly dangerous imported libraries/functions in Pickle-based files, anti-virus detection based on ClamAV \cite{clamav}, ModelScan \cite{modelscan}, which supports different AI model types including Pickle, Fickling \cite{Fickling}, and more.
Most of these scanners are employed on Hugging Face, and every file gets scan results from all the different scanners.
Unfortunately, these scanners are insufficient; they may miss serialization attacks \cite{pickle-attack}, and they have a very high false-positive rate \cite{malhug}.
Moreover, works such as Hide-and-Seek \cite{hide_and_seek} demonstrate how clever techniques can bypass these scanners by constructing more than 20 malicious Pickle-based models that do not get classified correctly by any of these scanners.

MalHug \cite{malhug} presented an investigation of different threats that exist on PTM sharing hubs, including the PDRE, and they introduced an LLM-based classifier approach to detecting Pickle-based file attacks; however, they did not publish their code or evaluation results.

PickleBall \cite{PickleBall2025} investigated the PDRE threat on HuggingFace, concluding that Pickle is still consistently used by AI model users, and introducing a policy-based safe Pickle loader similar to the restricted Unpickler. Additionally, they provide a curated dataset of benign and malicious Pickle-based files. The added benefit of the method is the automatic generation of blacklist/whitelist policy based on supplied benign models of a certain AI library (e.g., Conch, diffusers, etc.). We compare our method to PickleBall by using their supplied detection methodology.

Other works investigate AI model deserialization attacks in other serialization formats besides Pickle; for example, TensorAbuse \cite{tf_abuse} analyses the ability to create malicious TensorFlow models, and they introduce evasive attacking capabilities.


\section{Conclusion}
This work presented \textbf{SafePickle}, a machine-learning framework for detecting malicious Pickle-based payloads through opcode-level static analysis. Across a comprehensive set of experiments that included curated training data, external datasets, multi-format adversarial attacks, and operational malware, we demonstrated that SafePickle consistently demonstrates superior true-positive and true-negative rates, higher F1-scores, and significantly faster runtime performance compared to existing SOTA scanners.

Where traditional tools fail under compression, archiving, format transformation, or wrapper obfuscation, SafePickle maintains reliable detection by learning intrinsic structural properties of Pickle execution. The framework generalizes across domains, identifies previously unseen threats, and delivers millisecond-level runtimes suitable for modern deployment pipelines.

Overall, the findings suggest that \textbf{ML-driven static analysis represents a viable and highly effective strategy for securing Python serialization ecosystems}. SafePickle provides a scalable foundation for future research toward comprehensive, interpretable, and resilient detection of serialization-based threats in machine-learning environments.

\section{Open Science \& Data Availability}
We firmly support open and transparent science and the sharing of research artifacts.
We publish our curated dataset at \cite{ourdataset}.
Unfortunately, we can not publish the code of our experiments since it is proprietary.

However, we discuss the methodology, datasets, and experimental setup we used to an extent that allows complete recreation of our experimental results and analysis.
Our solution is comprised of known feature engineering, and ML techniques using the well-known Scikit-learn library, therefore, the technical replication of our method is very simple. In fact, part of the aim of our method was to give a simpler and more generic solution to Pickle scanning.

\cleardoublepage

\bibliographystyle{plain}
\bibliography{main}

\begin{thebibliography}{10}

\bibitem{modelscan}
Protect AI.
\newblock Modelscan: Protection against model serialization attacks.
\newblock Accessed: 2023-05-01.

\bibitem{S4VEtheD4TE}
ANONYMOUSLGD.
\newblock S4vethed4te ransomware written in c\# using windows forms.
\newblock Accessed: 2023-05-01.

\bibitem{mitre_atlas}
MITRE ATLAS.
\newblock Mitre atlas, 2024.
\newblock Accessed: 2024-05-01.

\bibitem{ourdataset}
Article Authors.
\newblock {SafePickle Data Repository}, 2025.
\newblock \url{https://huggingface.co/HFscanner1231}.

\bibitem{clamav}
{Cisco Talos}.
\newblock Clamav: Open source antivirus engine.
\newblock \url{https://github.com/Cisco-Talos/clamav}.
\newblock Version 1.5, commit c73755d. Accessed: 2023-05-01.

\bibitem{jfrog_ai_model}
David Cohen.
\newblock Data scientists targeted by malicious hugging face ml models with silent backdoor, 2024.
\newblock Accessed: 2024-05-01.

\bibitem{davis2023reusing}
James~C Davis, Purvish Jajal, Wenxin Jiang, Taylor~R Schorlemmer, Nicholas Synovic, and George~K Thiruvathukal.
\newblock Reusing deep learning models: Challenges and directions in software engineering.
\newblock In {\em 2023 IEEE John Vincent Atanasoff International Symposium on Modern Computing (JVA)}, pages 17--30. IEEE, 2023.

\bibitem{Huggingfacecite}
Hugging Face.
\newblock Hugging face model zoo.
\newblock Accessed: 2023-05-01.

\bibitem{Safetensors}
Hugging Face.
\newblock Safetensors.
\newblock Accessed: 2024-08-01.

\bibitem{danigil_stega_1}
Daniel Gilkarov and Ran Dubin.
\newblock Steganalysis of ai models lsb attacks.
\newblock {\em IEEE Transactions on Information Forensics and Security}, 19:4767--4779, 2024.

\bibitem{gilkarov2025neupermdisruptingmalwarehidden}
Daniel Gilkarov and Ran Dubin.
\newblock Neuperm: Disrupting malware hidden in neural network parameters by leveraging permutation symmetry, 2025.

\bibitem{github}
GitHub.
\newblock {GitHub: Software Sharing Hub}, 2025.
\newblock Accessed: 2025-06-05.

\bibitem{AWS_AISEC_PICKLE}
Nur Gucu and AWS~Security Matthew~Schwartz.
\newblock Enhancing cloud security in ai/ml: The little pickle story, 2025.
\newblock Accessed: 2025-05-05.

\bibitem{maleficnet2}
Dorjan Hitaj, Giulio Pagnotta, Fabio~De Gaspari, Sediola Ruko, Briland Hitaj, Luigi~V. Mancini, and Fernando Perez-Cruz.
\newblock Do you trust your model? emerging malware threats in the deep learning ecosystem, 2024.

\bibitem{maleficnet1}
Dorjan Hitaj, Giulio Pagnotta, Briland Hitaj, Luigi~V. Mancini, and Fernando Perez-Cruz.
\newblock Maleficnet: Hiding malware into deep neural networks using spread-spectrum channel coding.
\newblock In Vijayalakshmi Atluri, Roberto Di~Pietro, Christian~D. Jensen, and Weizhi Meng, editors, {\em Computer Security -- ESORICS 2022}, pages 425--444, Cham, 2022. Springer Nature Switzerland.

\bibitem{lora}
Edward~J. Hu, Yelong Shen, Phillip Wallis, Zeyuan Allen-Zhu, Yuanzhi Li, Shean Wang, Lu~Wang, and Weizhu Chen.
\newblock Lora: Low-rank adaptation of large language models, 2021.

\bibitem{huang2022pain}
Nan-Jung Huang, Chih-Jen Huang, and Shih-Kun Huang.
\newblock Pain pickle: Bypassing python restricted unpickler for automatic exploit generation.
\newblock In {\em 2022 IEEE 22nd International Conference on Software Quality, Reliability and Security (QRS)}, pages 1079--1090. IEEE, 2022.

\bibitem{jfrog_article1}
JFrog.
\newblock Data scientists targeted by malicious hugging face ml models with silent backdoor, 2024.

\bibitem{jiang2023empiricalstudypretrainedmodel}
Wenxin Jiang, Nicholas Synovic, Matt Hyatt, Taylor~R. Schorlemmer, Rohan Sethi, Yung-Hsiang Lu, George~K. Thiruvathukal, and James~C. Davis.
\newblock An empirical study of pre-trained model reuse in the hugging face deep learning model registry, 2023.

\bibitem{ptm_hf_survey}
Wenxin Jiang, Nicholas Synovic, Rohan Sethi, Aryan Indarapu, Matt Hyatt, Taylor~R. Schorlemmer, George~K. Thiruvathukal, and James~C. Davis.
\newblock An empirical study of artifacts and security risks in the pre-trained model supply chain.
\newblock In {\em Proceedings of the 2022 ACM Workshop on Software Supply Chain Offensive Research and Ecosystem Defenses}, SCORED'22, page 105–114, New York, NY, USA, 2022. Association for Computing Machinery.

\bibitem{pickleopcode}
Kaitai\_Project.
\newblock Pickle opcodes.
\newblock Accessed: 2023-05-01.

\bibitem{reversinglabs}
Reverse Engineer at~ReversingLabs Karlo~Zanki.
\newblock Malicious ml models discovered on hugging face platform.
\newblock \url{https://www.reversinglabs.com/blog/rl-identifies-malware-ml-model-hosted-on-hugging-face}, February 2025.

\bibitem{PickleBall2025}
Andreas~D. Kellas, Neophytos Christou, Wenxin Jiang, Penghui Li, Laurent Simon, Yaniv David, Vasileios~P. Kemerlis, James~C. Davis, and Junfeng Yang.
\newblock Pickleball: Secure deserialization of pickle-based machine learning models (extended report), 2025.

\bibitem{hf_token_leak}
Bar Lanyado.
\newblock +1500 huggingface api tokens were exposed, leaving millions of meta-llama, bloom, and pythia users vulnerable.
\newblock Accessed: 2025-05-05.

\bibitem{hide_and_seek}
Tong Liu, Guozhu Meng, Peng Zhou, Zizhuang Deng, Shuaiyin Yao, and Kai Chen.
\newblock The art of hide and seek: Making pickle-based model supply chain poisoning stealthy again, 2025.

\bibitem{picklescan}
Matthieu Maitre.
\newblock {Picklescan: Security scanner detecting Python Pickle files performing suspicious actions}, 2025.
\newblock Accessed: 2025-06-04.

\bibitem{hf_aijacking}
Nadav Noy.
\newblock Legit discovers "ai jacking" vulnerability in popular hugging face ai platform.
\newblock Accessed: 2025-05-05.

\bibitem{ohm2020backstabber}
Marc Ohm, Henrik Plate, Arnold Sykosch, and Michael Meier.
\newblock Backstabber’s knife collection: A review of open source software supply chain attacks.
\newblock In {\em International Conference on Detection of Intrusions and Malware, and Vulnerability Assessment}, pages 23--43. Springer, 2020.

\bibitem{onnx}
ONNX.
\newblock Open neural network exchange.
\newblock Accessed: 2023-05-01.

\bibitem{scikit-learn}
F.~Pedregosa, G.~Varoquaux, A.~Gramfort, V.~Michel, B.~Thirion, O.~Grisel, M.~Blondel, P.~Prettenhofer, R.~Weiss, V.~Dubourg, J.~Vanderplas, A.~Passos, D.~Cournapeau, M.~Brucher, M.~Perrot, and E.~Duchesnay.
\newblock Scikit-learn: Machine learning in {P}ython.
\newblock {\em Journal of Machine Learning Research}, 12:2825--2830, 2011.

\bibitem{pickle_standard}
Pickle.
\newblock Pickle- python object serialization.
\newblock Accessed: 2023-05-01.

\bibitem{PickleRestriction}
Pickle.
\newblock Restricting globals.
\newblock Accessed: 2024-09-15.

\bibitem{pickletools}
Python.
\newblock {pickletools}, 2025.
\newblock Accessed: 2025-06-04.

\bibitem{pytorchhub}
PyTorch.
\newblock Pytorch model sharing hub, 2022.
\newblock Accessed: 2022-01-15.

\bibitem{pytorch_serialization}
PyTorch.
\newblock Pytorch saving/loading, 2024.

\bibitem{Pickle_serialization_vul}
Marco Slaviero.
\newblock Sour pickles, a serialized exploitation guide in one part.
\newblock Accessed: 2023-05-01.

\bibitem{model-fickling}
Evan Sultanik.
\newblock Fickling is a decompiler, static analyzer, and bytecode rewriter for python pickle object serializations, 2021.
\newblock Accessed: 2022-01-15.

\bibitem{Fickling}
Evan Sultanik.
\newblock {Fickling pickle scanning}, 2021.
\newblock Accessed: 2022-12-19.

\bibitem{pickle-attack}
Evan Sultanik.
\newblock {Never a dill moment: Exploiting machine learning pickle files}, 2022.
\newblock Accessed: 2022-12-19.

\bibitem{wang2021evilmodel}
{Z. Wang}, {C. Liu}, and {X. Cui}.
\newblock Evilmodel: hiding malware inside of neural network models.
\newblock In {\em 2021 IEEE Symposium on Computers and Communications (ISCC)}, pages 1--7. IEEE, 2021.

\bibitem{wang2022evilmodel}
{Z. Wang}, {C. Liu}, {X. Cui}, {J. Yin}, and {X. Wang}.
\newblock Evilmodel 2.0: bringing neural network models into malware attacks.
\newblock {\em Computers \& Security}, 120:102807, 2022.

\bibitem{malhug}
Jian Zhao, Shenao Wang, Yanjie Zhao, Xinyi Hou, Kailong Wang, Peiming Gao, Yuanchao Zhang, Chen Wei, and Haoyu Wang.
\newblock Models are codes: Towards measuring malicious code poisoning attacks on pre-trained model hubs.
\newblock In {\em Proceedings of the 39th IEEE/ACM International Conference on Automated Software Engineering}, ASE ’24, page 2087–2098. ACM, October 2024.

\bibitem{dl_steganalysis_1}
Na~Zhao, Kejiang Chen, Chuan Qin, Yi~Yin, Weiming Zhang, and Nenghai Yu.
\newblock Calibration-based steganalysis for neural network steganography.
\newblock In {\em Proceedings of the 2023 ACM Workshop on Information Hiding and Multimedia Security}, pages 91--96, 2023.

\bibitem{tf_abuse}
Ruofan Zhu, Ganhao Chen, Wenbo Shen, Xiaofei Xie, and Rui Chang.
\newblock My model is malware to you: Transforming ai models into malware by abusing tensorflow apis.
\newblock In {\em 2025 IEEE Symposium on Security and Privacy (SP)}, pages 12--12. IEEE Computer Society, 2024.

\bibitem{zhu2024my}
Ruofan Zhu, Ganhao Chen, Wenbo Shen, Xiaofei Xie, and Rui Chang.
\newblock My model is malware to you: Transforming ai models into malware by abusing tensorflow apis.
\newblock In {\em 2025 IEEE Symposium on Security and Privacy (SP)}, pages 12--12. IEEE Computer Society, 2024.

\end{thebibliography}

\end{document}